\documentclass[a4paper,11pt]{article}
\usepackage{pos}

\usepackage{subfig}
\usepackage{float}

\numberwithin{equation}{section}
\numberwithin{figure}{section}
\numberwithin{table}{section}

\title{Gauge-Fixed Fourier Acceleration}
\author*{Ahmed Sheta} 
\notes{\note{This research was supported by the Exascale Computing Project (17-SC-20-SC), a collaborative effort of the US DOE’s Office of Science and National Nuclear Security Administration.}}
\author{Yidi Zhao}
\author{Norman H. Christ}

\affiliation{Department of Physics, Columbia University,\\
New York, NY 10027, USA}

\emailAdd{as5452@columbia.edu}
\emailAdd{yz3210@columbia.edu}
\emailAdd{nhc@phys.columbia.edu}

\abstract{For an asymptotically free theory, a promising strategy for eliminating Critical Slowing Down (CSD) is naïve Fourier acceleration. This requires the introduction of gauge-fixing into the action, in order to isolate the asymptotically decoupled Fourier modes. In this article, we present our approach and results from a gauge-fixed Fourier-accelerated hybrid Monte Carlo algorithm, using an action that softly fixes the gauge links to Landau gauge. We compare the autocorrelation times with those of the pure hybrid Monte Carlo algorithm. We work on a small-volume lattice at weak coupling.  We present preliminary results and obstacles from working with periodic boundary conditions, and then we present results from using fixed, equilibrated boundary links to avoid $\mathbb{Z}_3$ and other topological barriers and to anticipate applying a similar acceleration to many small cells in a large, physically-relevant lattice volume.}

\FullConference{%
 The 38th International Symposium on Lattice Field Theory, LATTICE2021
  26th-30th July, 2021
  Zoom/Gather@Massachusetts Institute of Technology
}

\begin{document}
\maketitle

\section{Introduction}

% One of the main obstacles to performing more accurate numerical lattice QCD calculations is Critical Slowing Down (CSD), which \textbf{?plagues/hinders} the hybrid Monte Carlo (HMC) algorithm (Ref needed?). 

As lattice QCD calculations are done at smaller lattice spacings in approaching the continuum limit ($a \to 0$), the frequencies affecting the simulation extend over a larger range to include more high-energy modes (with $\omega \propto 1/a$). These modes require a small molecular dynamics step size in order for their large forces to be accurately integrated. However, since the hybrid Monte Carlo (HMC) algorithm evolves all the modes with the same velocity, the more physical low-energy modes then require a large number of steps in order to detectably change from their old configurations. This problem slows down the generation of new gauge configurations by the HMC algorithm, and is usually referred to as Critical Slowing Down (CSD). CSD presents one of the main obstacles to performing more accurate numerical lattice QCD calculations at finer lattice spacings.

Because of QCD's asymptotic freedom, the majority of the gauge degrees of freedom enter the action quadratically in the continuum limit, mimicking a free field theory where the Fourier modes asymptotically decouple and perform simple harmonic motion. Hence, we introduced the Gauge-Fixed Fourier Acceleration (GFFA) in a previous article, an algorithm which attempts to utilize Fourier acceleration to eliminate CSD \cite{Yidi}.  In contrast to the HMC algorithm that uses the same mass for all the modes, GFFA uses a mode-dependent mass so that the low-frequency modes evolve at larger velocities than the high-frequency modes. In principle, this approach should eliminate the slow-down associated with the range of spacetime scales of Fourier frequencies, accelerating the simulation by a factor of $L$ for a given $L^4$ lattice. In this article, we briefly review the formulation we use of Fourier acceleration and the required gauge-fixing, and then we discuss the numerical results obtained by using the GFFA algorithm in generating gauge-configurations.

% A main challenge for numerical lattice QCD simulations is generating sufficiently many statistically independent gauge configurations in order to correctly estimate physical observables of interest. 

\section{Gauge-Fixing and Fourier Acceleration}

A barrier to applying naive Fourier acceleration is the gauge symmetry of the QCD action, which mixes the Fourier modes so they don't match the normal modes corresponding to the action. In order to overcome this obstacle, we introduce the following gauge-fixing term in the action
\begin{equation}
    \label{eq:GF term}
    S_{\mathrm{GF}} [U] = -\beta M^2 \sum_{x, \mu} \mathrm{Re} \left( tr [U_\mu (x)] \right),
\end{equation}
where $M$ is a parameter that controls how strongly the gauge is fixed. $S_{\mathrm{GF}}$ is minimized when all of the links are in Landau gauge, such that the effect of adding it to the action is to softly fix the gauge of the lattice such that configurations closer to Landau gauge are favored in the stochastic evolution. 

In order to preserve the expectations of gauge-invariant observables, it was shown in Refs. \cite{Yidi, FP1, FP2} that we also need to add the following compensating Fadev-Poppov term in the action
\begin{equation}
    S_{\mathrm{FP}}[U] = \ln \int dg e^{-S_{\mathrm{GF}}[U^g]},
\end{equation}
so that the action takes the form
\begin{equation}
    \label{eq:action}
    S[U] = S_{\mathrm{Wilson}}[U] + S_{\mathrm{GF}} [U] + S_{\mathrm{FP}} [U].
\end{equation}

The action enters the numerical simulation through the associated forces, which, for $S_{\mathrm{Wilson}}$ and $S_{\mathrm{GF}}$, are simple functions of the links that are easy to calculate. The Fadev-Poppov force, on the other hand, turns out to be the more complicated integral over gauge transformations,
\begin{equation}
    \frac{\partial S_{\mathrm{FP}}[U]}{\partial U_l} = \frac{\int dg \frac{\partial S_{\mathrm{GF}}[U^g]}{\partial U_l} e^{-S_{\mathrm{GF}}[U^g]}}{\int dg e^{-S_{\mathrm{GF}}[U^g]}} = \left\langle \frac{\partial S_{\mathrm{GF}}[U^g]}{\partial U_l} \right\rangle_g .
\end{equation}

We stochastically estimate this expecation as an inner Monte Carlo computation over gauge transformations with the weight factor $e^{-S_{\mathrm{GF}}[U^g]}$, such that
\begin{equation}
    \label{eq:innerMC}
    \frac{\partial S_{\mathrm{FP}}[U]}{\partial U_l} \approx \frac{1}{N} \sum_{i=1}^{N} \frac{\partial S_{\mathrm{GF}}[U^{g_i}]}{\partial U_l}.
\end{equation}

Likewise, the change in the action over the course of a trajectory, required for calculating $\Delta H$ in the accept-reject step, is
\begin{equation}
    \begin{aligned}
    S_{\mathrm{FP}}[U'] - S_{\mathrm{FP}}[U] &= \ln \frac{\int dg e^{-S_{\mathrm{GF}}[U^g]} \cdot e^{S_{\mathrm{GF}}[U^g] - S_{\mathrm{GF}}[U'^g]}}{\int dg e^{-S_{\mathrm{GF}}[U^g]}} \\
    &\approx \ln \frac{1}{N} \sum_{i=1}^{N} e^{S_{\mathrm{GF}}[U^{g_i}] - S_{\mathrm{GF}}[U'^{g_i}]}.
    \end{aligned}
\end{equation}

The expectation over gauge transformations in this case involves finding the average of an exponential, which is dominated by a small subset of the sample space where the exponential takes on extraordinarily large values compared to the typical value. In our numerical experiments, an accurate estimate of $S_{\mathrm{FP}}[U'] - S_{\mathrm{FP}}[U]$ required a huge number of samples and was computationally impractical, and hence we abandon the accept-reject step and allow for finite step size errors.

Finally, given the kinetic term of the molecular dynamics Hamiltonian 
\begin{equation}
    H_p = \sum_{k} tr \left[ P_\mu (-k) D^{\mu \nu} (k) P_\nu (k) \right],
\end{equation}
Fourier acceleration is achieved by choosing the mass term $D_{\mu \nu}$ to be the inverse of terms in the action quadratic in the gauge fields. For our gauge-fixing action in Eq. (\ref{eq:action}), $D_{\mu \nu}$ in the continuum limit has been worked out up to first order in Ref. \cite{Fachin} as
\begin{equation}
    \label{eq:mass}
    \begin{aligned}
    & D_{\mu \nu} (k) = \frac{1}{k^2} P_{\mu \nu} ^T (k) + \frac{1}{M^2} P_{\mu \nu} ^L (k), \\
    & P_{\mu \nu} ^T(k) = \delta_{\mu \nu} - \frac{k_\mu k_\nu}{k^2}, \\
    & P_{\mu \nu} ^L(k) = \frac{k_\mu k_\nu}{k^2}.
    \end{aligned}
\end{equation}

The numerical implementation of Fourier acceleration has been analyzed in Ref. \cite{Yidi}. We note in particular the introduction of the parameter $\epsilon$, which gives a finite nonzero mass to the cyclic modes.

\section{Numerical Results at Periodic Boundary Conditions}

% \subsection{Gauge Fourier Modes}
Because of the gauge-fixing, Fourier modes of the vector potential are expected to perfrom simple harmonic motion with known frequencies in the continuum limit. Using the regular HMC kinetic term (with $D_{\mu \nu}(k) = \delta_{\mu \nu}$), the frequencies of the transverse modes are $k$-dependent as follows
\begin{equation}
    \label{eq:freq_HMC}
    \omega_k = \sqrt{\frac{\beta}{6}} k,
\end{equation}
while using the Fourier accelerated kinetic term eliminates the $k$-dependence of the frequencies such that
\begin{equation}
    \label{eq:freq_GFFA}
    \omega_k = \sqrt{\frac{\beta}{6}}.
\end{equation}

\begin{figure}[H]%
    \centering
    \subfloat[HMC kinetic term] {{\includegraphics[width=0.5\textwidth] {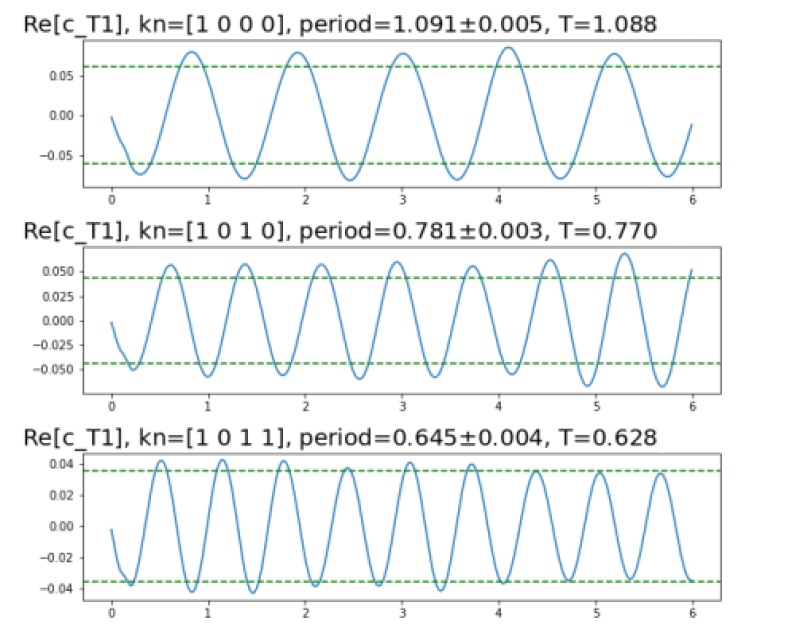} }}%
    \subfloat[Fourier accelerated kinetic term]
    {{\includegraphics[width=0.5\textwidth] {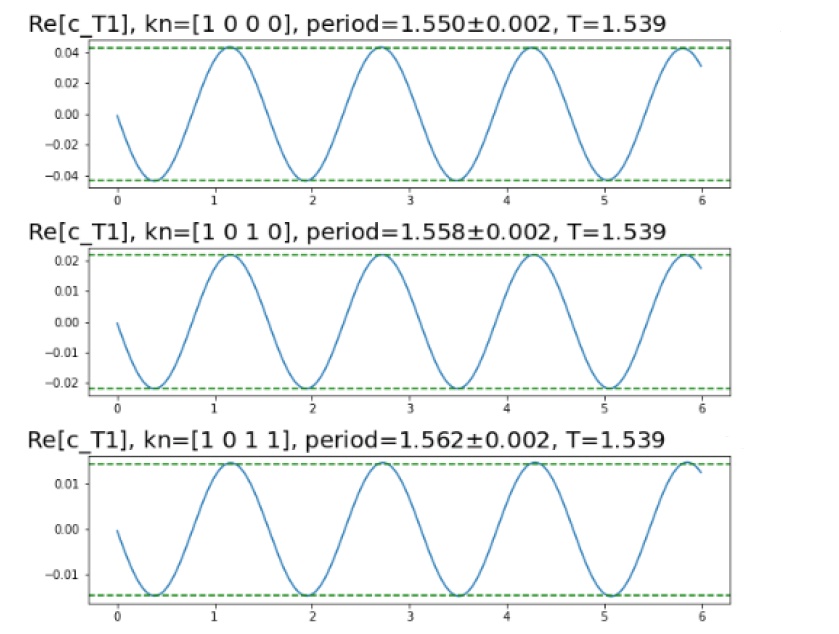} }}%
    \caption{Evolution of the real part of three different Fourier modes of the vector potential ($A = \ln U$) with respect to Monte Carlo time, using the gauge-fixing action. Figure (a) uses the HMC kinetic term ($D_{\mu \nu} = \delta_{\mu \nu}$), while figure (b) uses the Fourier accelerated kinetic term, with $D_{\mu \nu}$ given by Eq. (\ref{eq:mass}). These results come from a $4^4$ lattice at $\beta=100$.}%
    \label{fig:gauge modes}%
\end{figure}

In Figure \ref{fig:gauge modes}, we examine the numerical evolution of Fourier modes with respect to Monte Carlo time, using the gauge-fixing action of Eq. (\ref{eq:action}). The plots show the simple harmonic motion of the Fourier modes with frequencies described by Eqs. (\ref{eq:freq_HMC}) and (\ref{eq:freq_GFFA}); and in particular, the dependence of the oscillation frequency on the mode's spacetime scale vanishes if we use the Fourier accelerated kinetic term. 

% In Figure \ref{fig:gauge modes} below, we plot the numerical evolution as a function of Monte Carlo time of several Fourier modes using the gauge-fixing action.  In Figure \ref{fig:gauge modes}(a), we use the regular HMC kinetic term, observing the expected $k$-dependence of the frequencies described by equation Eq. (\ref{eq:freq_HMC}). On the other hand, we use the Fourier accelerated kinetic term in Figure \ref{fig:gauge modes}(b), and we observe the $k$-independence of oscillation frequencies. 

% \subsection{Autocorrelation Times}

In order to determine the speedup in the generation of independent gauge configurations, we compare integrated autocorrelation $\tau_{int}$ times for the plaquette and the Wilson flowed energy (flow time $\tau=4$) between the GFFA simulations and HMC simulations. We work in a small volume with periodic boundary conditions, at weak coupling $\beta=10$. The results, shown in Tables \ref{table:periodic_HMC} and \ref{table:periodic_GFFA}, indicate a factor of $\approx 5\times$ acceleration achieved by GFFA as opposed to HMC.

% The numerical results for the HMC runs are are shown in Table \ref{table:periodic_HMC}, and for the GFFA runs are shown in Table \ref{table:periodic_GFFA}. Comparing the integrated autocorrelation time for the plaquette, GFFA achieves a factor of $\approx 5\times$ acceleration over the HMC.

\begin{table}[H]
    \centering
    \begin{tabular}{|l|l|l|l|l|l|l|l|l|l|l|l|}
    \hline
    $\beta$ & $\tau_{traj}$   & steps & trajs & plaq          & plaq $\tau_{int}$ & E(4) $\tau_{int}$ &  Accpt  \\ \hline
    
    10 &       0.6 &       24 & 10030 & 0.783295(37) & 4.51(82) & 20.9(5.4) & 75\% \\ \hline
    
    10 &       1.0 &       50 & 10030 & 0.783289(26) & 3.41(30) & 19.0(4.4) & 77.6\% \\ \hline
    
    10 &       4.0 &       200 & 8533 & 0.783363(27) & 10.73(61) & 13.0(1.4) & 78\% \\ \hline
    \end{tabular}
\caption{HMC runs for an $8^4$ lattice with periodic boundary conditions, at $\beta=10$. Autocorrelation times are reported in molecular dynamics time units.}
\label{table:periodic_HMC}
\end{table}

\begin{table}[H]
    \centering
    \begin{tabular}{|l|l|l|l|l|l|l|l|l|l|l|l|l|l|}
    \hline
    $\beta$ & $\tau_{traj}$   & steps & trajs & plaq         & plaq $\tau_{int}$ & E(4) $\tau_{int}$ & M & MC \\ \hline
    
    10 &       0.6 &     24 &     1911 & 0.783347(45) & 1.26(25) & 3.6(1.0) & 3.0 & 200  \\ \hline
    
    10 &       1.0 &     30 &     5176 & 0.783207(16) & 0.653(31) & 18.6(8.3) & 3.0 & 40 \\ \hline
    
    10 &       0.6 &     24 &     3915 & 0.783404(28) & 1.059(81) & 5.4(1.3) & 5.0 & 200 \\ \hline
    \end{tabular}

\caption{GFFA runs for an $8^4$ lattice with periodic boundary conditions, at $\beta=10$. Autocorrelation times are reported in molecular dynamics time units.}
\label{table:periodic_GFFA}
\end{table}

% \subsection{Difficulties with Polyakov Phase}

Another observable of interest in the case of periodic boundary conditions is the phase of the average Polyakov loop along a fixed direction, or the Polyakov phase. The Wilson action enjoys a $\mathbb{Z}_3$ symmetry of the Polyakov phase, whereby the transformation $P \to e^{\frac{2 \pi}{3} i} P$ for all the Polyakov loops $P$ along the $\mu$ direction leaves the action invariant. While $S_{\mathrm{GF}}$ explicitly favors one of the three cube roots of identity, the $\mathbb{Z}_3$ symmetry should be preserved under our gauge-fixing action if we properly include the compensating Fadev-Poppov term.

\begin{figure}[H]
% \begin{subfigure}{.5\textwidth}
%   \centering
%   \includegraphics[width=.8\linewidth]{Z3_HMC.png}
%   \caption{1a}
%   \label{fig:sfig1}
% \end{subfigure}%
% \begin{subfigure}{.5\textwidth}
%   \centering
%   \includegraphics[width=.8\linewidth]{Z3_40.png}
%   \caption{1b}
%   \label{fig:sfig2}
% \end{subfigure}%

    \centering
    \subfloat[Gauge-fixed evolution; 40 inner MC samples ]
    {{\includegraphics[width=0.5\textwidth] {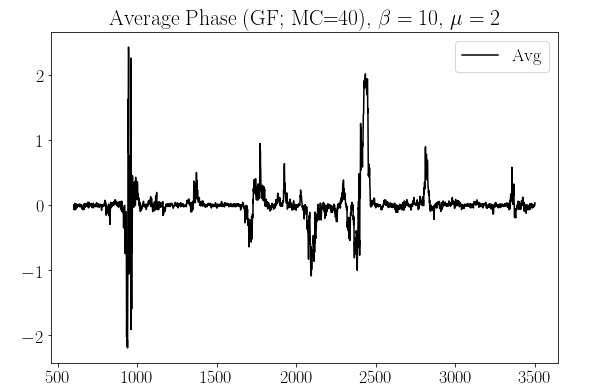} }} %
    \subfloat[Gauge-fixed evolution; 1000 inner MC samples ]
    {{\includegraphics[width=0.5\textwidth] {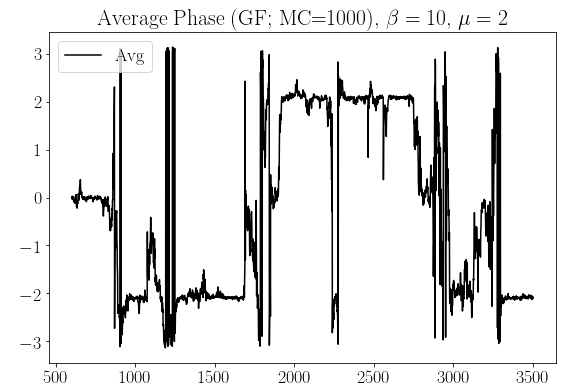} }}\\%
    \subfloat[HMC evolution] {{\includegraphics[width=0.5\textwidth] {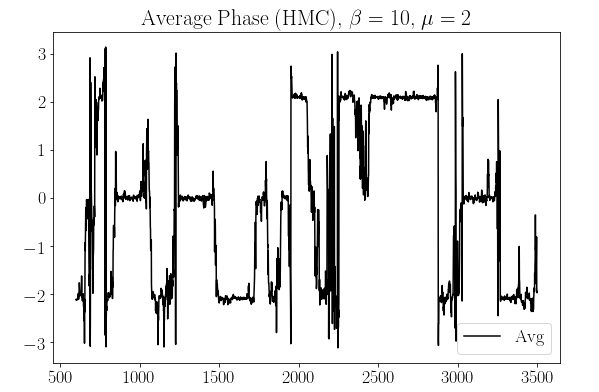} }}%
    \caption{Evolution of the phase of the average Polyakov loop in the $\mu=2$ direction, using the gauge-fixing action with 40 inner Monte Carlo samples (Figure a), gauge-fixing action with 1000 inner Monte Carlo samples (Figure b), and the pure the Wilson action (Figure c). Because of the inaccurate estimate of the Fadev-Poppov force, the evolution in Figure (a) is stuck at the $0$ phase and is unable to tunnel to the other two cube roots of the identity.}
    \label{fig:phases}
\end{figure}

Figure \ref{fig:phases} examines the evolution of the Polyakov phase in one direction. The HMC simulation shows occasional tunneling between the three cube roots of identity, properly respecting the underlying $\mathbb{Z}_3$ symmetry of the theory. The gauge-fixed simulations, on the other hand, require a very large number of Monte Carlo samples in estimating the Fadev-Poppov force in order to accurately compensate for the symmetry breaking of $S_{\mathrm{GF}}$ and properly tunnel. This presents computational difficulties in our experimental runs, and makes our interpretation of the reduction in autocorrelations seen between Tables \ref{table:periodic_HMC} and \ref{table:periodic_GFFA} uncertain. We overcome this obstacle by switching to a lattice with fixed boundary conditions, which is the setup we anticipate eventually working on, as will be explained in the next section. Changing the setup as such has the advantage of completely eliminating the $\mathbb{Z}_3$ symmetry, and is justified since these Polyakov phases are only nontrivial in high-temperature simulations, but they vanish in the confined regime that our approach aims to accelerate. We can therefore get by using a fewer number of inner Monte Carlo samples, but we must still check for correctness by demanding an accurate estimate of observables as compared to the HMC estimate, and ensuring more accurate convergence when using an increased number of Monte Carlo samples.

% Figure \ref{} show the evolution of the $\mu=2$ Polyakov phase with respect to trajectory number. Figure \ref{}(a) shows the standard HMC evolution (using the pure Wilson action), Figure \ref{}(b) shows the evolution using the gauge-fixing action with 40 Monte Carlo samples used to estimate the Fadev-Poppov force as per Eq. (\ref{eq:innerMC}), and Figure \ref{}(c) shows the evolution using the gauge-fixing action with 1000 Monte Carlo samples. 

\section{Fixed Boundary Conditions and Numerical Results}

% \subsection{Motivation and Setup}

The ultimate goal of our algorithm is to accelerate physically relevant calculations done on large lattices. Fourier acceleration, however, requires all the links in the lattice to be sufficiently close to identity. Even in the weak-coupling limit, a perturbative gauge (such as the Landau gauge of Eq. (\ref{eq:GF term})) is still required for the Fourier accelerated kinetic term to match the oscillation modes of the system. If the lattice is large enough to include nonperturbative QCD effects ($L \gtrsim 1/\Lambda_{QCD}$), finding a perturbative gauge for the entire lattice might be unattainable. 

\begin{figure}[H]%
    \centering
    \includegraphics[width=0.3\textwidth] {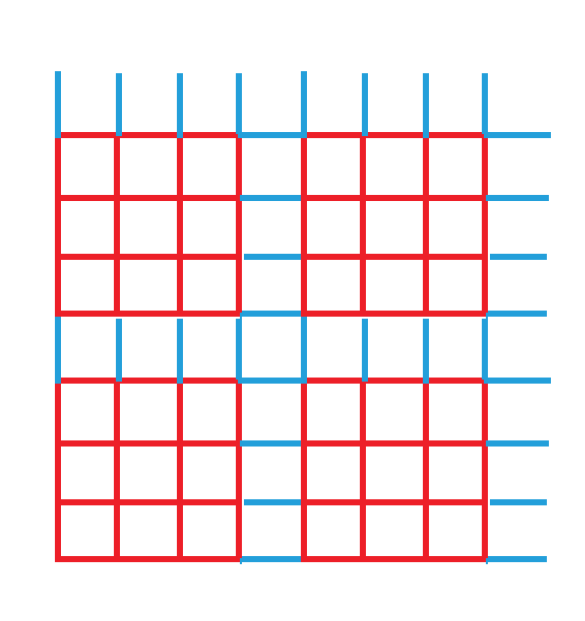}
    \caption{Strategy for accelerating calculations on physically large lattices. The lattice is divided into smaller cells (red links), which are independently evolved using the gauge-fixing, Fourier accelerated Hamiltonian of Eq. (\ref{eq:hamiltonian}), while the boundary links joining the cells (blue links) are held fixed.}%
    \label{fig:large vol}%
\end{figure}

Therefore, in order to accelerate simulations on large lattices, we divide the lattice into many cells which are small enough to be perturbative, as depicted in Figure \ref{fig:large vol}. Then, we evolve each of the cells independently, following a checkerboard scheme, using the gauge-fixing action of Eq. (\ref{eq:action}) and the Fourier accelerated kinetic term of Eq. (\ref{eq:mass}). If we start with the entire lattice equilibriated with respect to the Wilson action, then, for each cell, we softly fix into Landau gauge by sampling a gauge transformation $g(x)$ distributed according to the weight factor $e^{-S_{\mathrm{GF}}[U \in C] - S_{\mathrm{FP}}[U \in C]}$, where $U \in C$ are the links in the cell (the red links in Figure \ref{fig:large vol}). Applying $g(x)$ to the sites in the cell fixes the links of the cell to the perturbative Landau gauge and squeezes any nonperturbative effects into its exterior boundary links (the blue links in Figure \ref{fig:large vol}). Then, with boundary links held fixed, we evolve the cell according to the gauge-fixing, Fourier accelerated Hamiltonian
\begin{equation}
    \label{eq:hamiltonian}
    H = \sum_{k} tr \left[ P_\mu (-k) D^{\mu \nu} (k) P_\nu (k) \right] + S_{\mathrm{Wilson}} [U \in C] + S_{\mathrm{GF}} [U \in C] + S_{\mathrm{FP}} [U \in C],
\end{equation}
where $S_{\mathrm{Wilson}}[U \in C]$ includes every plaquette with any of its edges being one of the links in the cell, and $D_{\mu \nu}$ is the Fourier accelerated kinetic term in Eq. (\ref{eq:mass}), the Fourier modes being those of the cell. To ensure that all the degrees of freedom are updated, we occasionally move around the red cells to include the blue fixed boundary links as part of their evolving links.

% \subsection{Autocorrelation Times}

\begin{figure}
    \centering
    \subfloat[HMC run with $\tau_{traj} = 0.5$ and 2011 trajectories.]
    {{\includegraphics[width=0.5\textwidth]{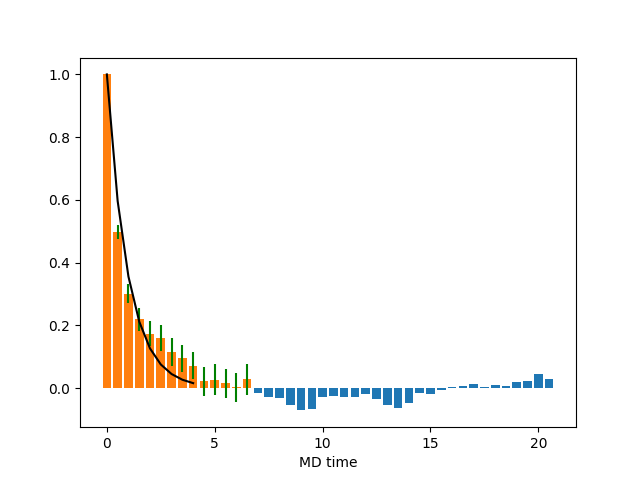} }} %
    \subfloat[GFFA run with $\tau_{traj} = 0.7$ and 1296 trajectories.]
    {{\includegraphics[width=0.5\textwidth]{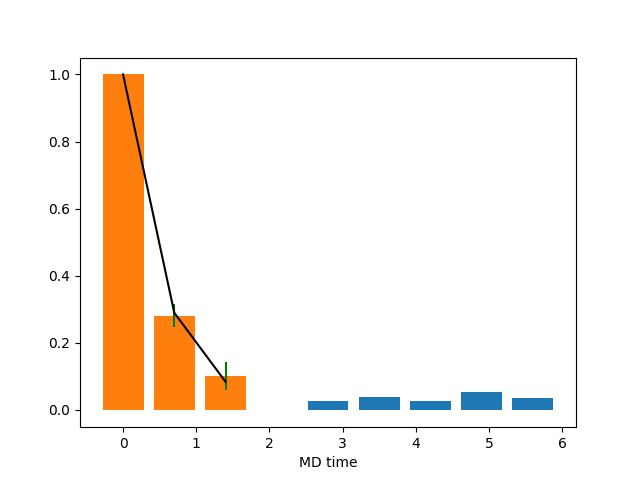} }}\\%
    \caption{Autcorrelation function of the plaquette on a $10^4$ lattice with fixed boundary conditions. The plots show a reduced autocorrelation in the GFFA case.}
    \label{fig:acf}
\end{figure}

We compare the autocorrelation function and exponential autocorrelation time $\tau_{exp}$ for the plaquette between HMC runs and GFFA runs in Figure \ref{fig:acf} and Tables \ref{table:fixed HMC 8}, \ref{table:fixed GFFA 8}, \ref{table:fixed HMC 12}, and \ref{table:fixed GFFA 12}. The exponential autocorrelation time is obtained from fitting the autocorrelation function to $f(t) = e^{-t/\tau_{exp}}$. In these experiments, we run the simulations on a lattice with fixed boundary conditions that are equilibriated with respect to Wilson action, equivalent to one of the cells in Figure \ref{fig:large vol}. 

Tables \ref{table:fixed HMC 12} and \ref{table:fixed GFFA 12} indicate a factor of $\approx 2 \times$ acceleration of the GFFA over the HMC algorithm. Additionally, if we compare the effect of increasing the size of the lattice from $6^4$ to $10^4$ on autocorrelations in the HMC and the GFFA runs, we observe the HMC attaining larger autocorrelations as opposed to the scale-independent GFFA that retains the same efficiency, consistent with a speedup proportional to $L$. This foreshadows an enhanced acceleration factor for the GFFA when working on larger lattices - simulations on which are currently being studied.

\begin{table}
    \centering
    \begin{tabular}{|l|l|l|l|l|l|l|l|l|l|l|l|}
    \hline
    $\beta$ & $\tau_{traj}$   & steps & trajs & plaq          & plaq $\tau_{exp}$ &  Accpt  \\ \hline
    
    10 &       0.5 &       48 & 9431 & 0.783318(56) & 0.810(45) & 97.2\% \\ \hline
    
    10 &       1.0 &       96 & 8481 & 0.783313(11) & 1.540(87) & 96.6\% \\ \hline
    \end{tabular}
    \caption{HMC runs for a $6^4$ lattice with fixed, equilibriated boundary conditions at $\beta=10$. Autocorrelation times are reported in molecular dynamics time units.}
    \label{table:fixed HMC 8}
\end{table}

\begin{table}
    \centering
    \begin{tabular}{|l|l|l|l|l|l|l|l|l|l|l|l|l|l|}
    \hline
    $\beta$ & $\tau_{traj}$   & steps & trajs & plaq         & plaq $\tau_{exp}$ & M & MC   \\ \hline
    
    10 &       0.7 &     48 &     2142 & 0.783111(14) & 0.581(47) & 3.0 & 200 \\ \hline
    \end{tabular}
    \caption{GFFA run for a $6^4$ lattice with fixed, equilibriated boundary conditions at $\beta=10$. Autocorrelation times are reported in molecular dynamics time units.}
    \label{table:fixed GFFA 8}
\end{table}

\begin{table}
    \centering
    \begin{tabular}{|l|l|l|l|l|l|l|l|l|l|l|l|}
    \hline
    $\beta$ & $\tau_{traj}$   & steps & trajs & plaq          & plaq $\tau_{exp}$ &  Accpt  \\ \hline
    10 &       0.5 &       48 & 2011 & 0.783395(13) & 0.97(14) & 93.3\% \\ \hline
    
    10 &       1.0 &       96 & 1807 & 0.783415(14) & 1.97(30) & 91.5\% \\ \hline
    \end{tabular}
    \caption{HMC runs for a $10^4$ lattice with fixed, equilibriated boundary conditions at $\beta=10$. Autocorrelation times are reported in molecular dynamics time units.}
    \label{table:fixed HMC 12}
\end{table}

\begin{table}
    \centering
    \begin{tabular}{|l|l|l|l|l|l|l|l|l|l|l|l|l|l|}
    \hline
    $\beta$ & $\tau_{traj}$   & steps & trajs & plaq         & plaq $\tau_{exp}$ & M & MC   \\ \hline
    10 &       0.7 &     60 &     1296 & 0.783101(10) & 0.569(58) & 3.0 & 200 \\ \hline
    \end{tabular}
    \caption{GFFA run for a $10^4$ lattice with fixed, equilibriated boundary conditions at $\beta=10$. Autocorrelation times are reported in molecular dynamics time units.}
    \label{table:fixed GFFA 12}
\end{table}

\section{Conclusion}
In this article, we reviewed the basic formulation of gauge-fixing and Fourier acceleration presented in Ref. \cite{Yidi}, and we presented numerical results from simulations on lattices with periodic boundary conditions. The $\mathbb{Z}_3$ symmetry presents a big computational barrier by requiring a huge number of inner Monte Carlo samples to accurately simulate, so we switch to a lattice with fixed boundary conditions in anticipation of applying GFFA to physically large lattices. Numerical results on lattices with fixed, equilibriated boundary conditions are presented as well, indicating a factor of $\approx 2\times$ acceleration in the plaquette observable. Analyzing appropriate observables that probe modes with longer spacetime scales might reveal an enhanced acceleration over the HMC algorithm. It also remains to run simulations on larger lattices that are still perturbative ($32^4$ at $\beta=10$ for example), and to check for stronger accelerations corresponding to the increased range of spacetime scales.

\end{document}